\documentstyle[aps,prd, preprint]{revtex}
\begin{document}

\title{Effects of geometry and topology on some quantum mechanical systems }
\author{Geusa de A. Marques and Valdir B. Bezerra.}
\maketitle

\begin{center}
\textit{{Departamento de F\'{\i}sica-CCEN, Universidade Federal da
Para\'{\i}ba,}}

\textit{{Caixa Postal 5008, 58051-970, Jo\~{a}o Pessoa, Pb, Brazil.}}
\end{center}

\begin{abstract}

We study the behaviour of a nonrelativistic quantum particle interacting
with different potentials in the spacetimes of topological defects. We find
the energy spectra and show how they differ from their free-space values.

\end{abstract}
\vskip 3.0 cm \centerline{\bf{I. Introduction} }

The study of quantum systems under the influence of a gravitational field
has been an exciting research field. It has been known that the energy
 eigenstates of an atom which interacts with a gravitational field are
 affected by the local spacetime curvature\cite{2,1}. As a result of this
 interaction, an observer at rest with respect
to the atom would see a change in its spectrum, which would depend on the
 features of the spacetime. The problem of
finding these shifts\cite{3} of the energy levels under the influence of a
gravitational field is of considerable theoretical interest as well as
possible observational. These shifts in the energy spectrum due to the
gravitational field are different from the ones produced by the
electromagnetic field present, for example, near white dwarfs and neutron
starts. In fact, it was already shown that in the Schwarzschild geometry,
the level spacing of the gravitational effect is different from that of the
well-known first order Stark and Zeeman effects, and therefore, in
principle, it would be possible to separate the electromagnetic and
gravitational perturbations of the spectrum\cite{3}. Other
investigations concerning this interesting subject is
to use loosely bound Rydberg atoms in
curved spacetime to study the gravitational shift in the
 energy spectrum\cite
{4}.

The first experiment which showed the gravitational effect on a wave
function was performed by Colella, Overhauser and
Werner\cite{5a} by measure of the quantum mechanical phase difference of
two neutron beams
induced by a gravitational field. Another gravitational
 effect that appears
in quantum interference due to a gravitational field is the
 neutrino oscillations\cite{6a}.

The general theory of relativity, as a metric theory, predict that
gravitation is manifested as curvature of spacetime. This curvature is
characterized by the Riemann tensor $R_{\beta \gamma \delta }^{\alpha }$.

It is of interest to know how the curvature of spacetime at the position of
the atom affects its spectrum. On the other hand, we know that there are
connections between topological properties of the space and local physical
laws in such a way that the local intrinsic geometry of the space is not
sufficient to describe completely the physics of a given system. Therefore,
it is also important to investigate the role played by a nontrivial
topology, for example, on a quantum system. As examples of these
investigations we can mention the calculation of the topological scattering
amplitude in the context of quantum mechanics on a cone\cite{5} and the
interaction of a quantum system with conical
singularities\cite{6,6b}. Therefore, an atom placed
 in a gravitational field is no longer exclusively influenced by
its interaction with the local curvature; the topology plays a role.

Then, the problem of finding how the energy spectrum of an atom placed
in a gravitational field is perturbed by this background has to take into
account the geometrical structure and topological features
 of the spacetime under
consideration. In other words, the dynamic of quantum systems
 is determined by the
curvature and also by the topology of the
background spacetime.

According to standard quantum mechanics, the motion of a charged particle
can be influenced by electromagnetic fields in regions from which the
particle is rigorously excluded\cite{10a}. In this region the
electromagnetic field vanishes and just inside a thin flux tube
 it does not vanish. This phenomenon has come to be called
Aharonov-Bohm effect after a seminal paper by Aharonov and
Bohm\cite{10a}. It was shown that in the quantum scattering
in accordance with the Aharonov-Bohm effects problem this
 background leads to
a non-trivial scattering, which was already confirmed
experimentally\cite{11a}.

The analogue of the electromagnetic Aharonov-Bohm effect
 set up in the gravitational framework is the
background spacetime of a cosmic string\cite{7,17} in
which the geometry is
flat everywhere apart from a symmetry axis. In this
scenario we also have a gravitational Aharonov-Bohm
effect for bound states\cite{6,19a}

Cosmic strings\cite{7,17} and monopoles\cite{8} are exotic
topological defects%
\cite{9} which may have been formed at phase transitions
 in the very early
history of the Universe. Up to the moment no direct
observational evidence
of their existence has been found, but the richness
of the new ideas they
brought along to general relativity seems to justify
the interest in the
study of these structures.

The gravitational field of a cosmic string is quite remarkable; a particle
placed at rest around a straight, infinite, static string will not be
attracted to it; there is no local gravity. The spacetime around a cosmic
string is locally flat but not globally. The external gravitational field
due to a cosmic string may be approximately described by a commonly called
conical geometry. Due to this conical geometry a cosmic string can induce
several effects like, for example, gravitational lensing \cite{10}, pair
production\cite{11}, electrostatic self-force\cite{12} on an electric charge
at rest, bremsstr\"{a}hlung radiation\cite{13} and the so-called
gravitational Aharonov-Bohm effect\cite{14}.

The spacetime of a point global monopole has also some unusual properties.
It possesses a deficit solid angle $\Delta =32\pi ^{2}G\eta ^{2}$, $\eta $
being the energy scale of symmetry breaking. Test particles in this
spacetime experiences a topological scattering by an angle $\pi \frac{\Delta
}{2}$ irrespective of their velocity and their impact parameter. The effects
produced by the point global monopole are due to the deficit solid angle
which determines the curvature and the topological features of this
spacetime.

In this paper we shall study the energy shift associated with a
non-relativistic quantum particle interacting with an harmonic oscillator
and a Coulomb potential with these systems placed in the background
spacetime of a cosmic string, and determine how the nontrivial topology of
this background spacetime perturbs the energy spectra. The influence of the
conical geometry on the energy eigenvalues manifests as a kind of
gravitational Aharonov-Bohm effect for bound states\cite{6,19a}, whose
analogue in the electromagnetic case shows that\cite{20a} that the bound
state energy dependents on the external magnetic flux in a region from which
the electron is excluded.

In the case of a point global monopole we are concerned with the similar
proposal. We will investigate the interactions of a non-relativistic quantum
particle with the Kratzer\cite{15} and the Morse
potentials\cite{16} placed in the background spacetime of a global
 monopole. In this case we also determine the shifts
in the energy levels.

These modifications in the energy spectra as compared to the simplest
situation of empty flat Minkowski spacetime, could be used, in principle, as
a probe of the presence of these defects in the cosmos. The possibility of
using this effects for such a purpose is interesting because at present our
observational knowledge of the existence of such defects is quite limited
and this effect offers one more possibility to detect these topological
defects.

To this end let us consider that a non-relativistic
particle living a curved spacetime is described by the Schr\"{o}dinger
equation which should take the form
\begin{equation}
i\hbar \frac{\partial \psi }{\partial t}=-\frac{\hbar ^{2}}{2\mu }\nabla
_{LB}^{2}\psi +V({\bf r}),  \label{a1}
\end{equation}
where $\nabla _{LB}^{2}$ is the Laplace-Beltrami operator the covariant
version of the Laplacian given by $\nabla _{LB}^{2}=g^{-\frac{1}{2}}\partial
_{i}\left( g^{ij}g^{\frac{1}{2}}\partial _{j}\right) $,
 with $i,j=1,2,3;$ $%
g=\det \left( g_{ij}\right)$ stands for the determinant of
 the metric $ g_{ij}$; $\mu $ is the mass of the particle
and $V({\bf r})$ is
an external potential. Throughout this paper we will use
units in which $c=1$%
.

\vskip 2.0 cm

\centerline{\bf{II. Harmonic oscillator in the spacetime of a cosmic string}}

The line element corresponding to the cosmic string spacetime is given in
cylindrical coordinates by\cite{17}

\begin{equation}
ds^2=-dt^2+d\rho ^2+\alpha ^2\rho ^2d\theta ^2+dz^2,  \label{1}
\end{equation}
where $\rho \geq 0$ and $0\leq \varphi \leq 2\pi $, the parameter $\alpha
=1-4G\bar{\mu}$ runs in the interval $\left[ 0,1\right] $, $\bar{\mu}$ being
the linear mass density of the cosmic string. The string is situated on the $%
z$-axis. In the special case $\alpha =1$ we obtain the Minkowski space in
cylindrical coordinates. This metric has a cone-like singularity at $\rho =0$%
. In other words, the curvature tensor of the metric (\ref{1}), considered
as a distribution, is of the form
\begin{equation}
R_{\text{ }12}^{12}=2\pi \frac{\alpha -1}\alpha \delta ^2(\rho ),  \label{a2}
\end{equation}
where $\delta ^2(\rho )$ is the two-dimensional Dirac $\delta $-function.

Now, let us consider the Schr\"{o}dinger equation in the metric (\ref{1})
which is given by
\begin{equation}
-\frac{\hbar ^2}{2\mu }\left[ \partial _\rho ^2+\frac 1\rho \partial _\rho +%
\frac 1{\alpha ^2\rho ^2}\partial _\theta ^2+\partial _z^2\right] \psi
(t,\rho ,\theta ,z)+V(\rho ,z)\psi (t,\rho ,\theta ,z)=i\hbar \frac \partial
{\partial t}\psi (t,\rho ,\theta ,z),  \label{a3}
\end{equation}
where $V(\rho ,z)$ is the potential energy corresponding to a
three-dimensional harmonic oscillator which is assumed to be
\begin{equation}
V(\rho ,z)=\frac 12\mu w^2\left( \rho ^2+z^2\right) .  \label{2}
\end{equation}
In what follows, we shall solve the above Schr\"{o}dinger
 equation with the interaction potential given
by Eq. (\ref{2}). To do this let us use the method
of separation of variables by searching for solutions of
the form
\begin{equation}
\psi (t,\rho ,\theta ,z)=\frac 1{\sqrt{2\pi }}e^{-iEt+im\theta }R(\rho )Z(z),
\label{a4}
\end{equation}
with ${m}$ an integer.
Using Eq. (\ref{a4}), Eq. (\ref{a3}) yields to the following
 ordinary differential equations for $R(\rho
)$ and $Z(z)$

\begin{equation}
-\frac{\hbar ^{2}}{2\mu }\left[ \frac{1}{R(\rho )}\frac{\partial ^{2}R(\rho )%
}{\partial \rho ^{2}}+\frac{1}{R(\rho )\rho }\frac{\partial R(\rho )}{%
\partial \rho }-\frac{m^{2}}{\alpha ^{2}\rho ^{2}}\right] +\frac{1}{2}\mu
w^{2}\rho ^{2}=\epsilon  \label{9}
\end{equation}
and
\begin{equation}
-\frac{\hbar ^{2}}{2\mu Z(z)}\frac{\partial ^{2}Z(z)}{\partial z^{2}}+\frac{1%
}{2}\mu w^{2}z^{2}=\varepsilon _{z},  \label{10}
\end{equation}
where $\epsilon $ is a separation constant and such that
\begin{equation}
\epsilon +\varepsilon _{z}=E.  \label{11}
\end{equation}
Equation (\ref{10}) is the Schr\"{o}dinger equation for a particle in the
presence of one-dimensional harmonic oscillator potential, and then we have
the well-known results
\begin{equation}
\varepsilon _{z}=\left( n_{z}+\frac{1}{2}\right) \hbar w;\text{ }%
n_{z}=0,1,2,...  \label{12}
\end{equation}
with
\begin{equation}
Z(z)=2^{-\frac{n_{z}}{2}}\left( n_{z}!\right) ^{-\frac{1}{2}}\left( \frac{%
\mu w}{\hbar \pi }\right) ^{\frac{1}{4}}e^{-\frac{\mu w}{2\hbar }%
z^{2}}H_{n_{z}}\left( \sqrt{\frac{\mu w}{\hbar }}z\right) ,  \label{a5}
\end{equation}
where $H_{n_{z}}$ denotes a Hermite Polynomial of degree $n_z $

Solving Eq. (\ref{9}) we get
\begin{equation}
R(\rho )=\exp \left( -\frac \tau 2\rho ^2\right) \rho ^{\frac{\left|
m\right| }\alpha }F(\rho ),  \label{13}
\end{equation}
where $\tau =\frac{mw}\hbar $ and
\begin{equation}
F(\rho )=_1F_1\left( \frac 12+\frac{\left| m\right| }{2\alpha }-\frac{\mu
\epsilon }{2\hbar ^2\tau },\frac A2;\tau \rho ^2\right)  \label{15}
\end{equation}
is the degenerate hypergeometric function, with $A=1+2\frac{\left| m\right| }%
\alpha $.

In order to have normalizable wavefunction, the series in Eq. (\ref{15})
must be a polynomial of degree $n_{\rho }$, and therefore
\begin{equation}
\frac{1}{2}+\frac{\left| m\right| }{2\alpha }-\frac{\mu \epsilon }{2\hbar
^{2}\tau }=-n_{\rho };\text{ }n_{\rho }=0,1,2,....  \label{17}
\end{equation}
With this condition, we obtain the following result
\begin{equation}
\epsilon =\hbar w\left( 1+\frac{\left| m\right| }{\alpha }+2n_{\rho }\right) .
\label{18}
\end{equation}
If we substitute Eqs. (\ref{18}) and (\ref{12}) into (\ref{11}) we get,
finally, the energy eigenvalues
\begin{equation}
E_{N}=\hbar w\left( N+\frac{\left| m\right| }{\alpha }+\frac{3}{2}\right) ,
\label{19}
\end{equation}
where $N=2n_{\rho }+n_{z}.$

 From this expression we can conclude that the
presence of the cosmic string breaks the degeneracy of the energy levels.

The complete eigenfunctions are then given by
\begin{eqnarray}
\psi (t,\rho ,\theta ,z) &=&C_{Nm}e^{-iE_N\frac t\hbar }e^{-\frac \tau 2\rho
^2}\rho ^{\frac{\left| m\right| }\alpha }F_1\left( \frac 12+\frac{\left|
m\right| }{2\alpha }-\frac{\mu \epsilon }{2\hbar ^2\tau },\frac A2;\tau \rho
^2\right)  \nonumber \\
&&e^{im\theta }2^{-\frac{n_z}2}\left( n_z!\right) ^{-\frac 12}\left( \frac{%
\mu w}{\hbar \pi }\right) ^{\frac 14}e^{-\frac{\mu w}{2\hbar }%
z^2}H_{n_z}\left( \sqrt{\frac{\mu w}\hbar }z\right) ,  \label{b2}
\end{eqnarray}
where $C_{Nm}$ is a normalization constant.

The results corresponding to the energy eigenvalues and
 eigenfunctions given by Eqs. (\ref{19}) and (\ref{b2}),
 respectively, recover the ones for the flat spacetime
 in the limit $\alpha \rightarrow 1$ as it should be.

As an estimation of the effect of the cosmic string on the energy shift, let
us consider $\alpha \cong 0.999999,$ which corresponds to GUT cosmic
strings. In this case the shift in the energy spectrum between the first two
levels increases of about $10^{-5}\%$ as compared with
the flat Minkowski spacetime case.\vskip 2.0 cm

\centerline{\bf{III. Coulomb potential in the spacetime of a cosmic string}}

Now, let us determine the energy eigenvalues of a particle in the presence of
a Coulomb potential in the spacetime of a cosmic string. To do this let us
consider the exterior metric of an infinitely long straight and static
string in spherical coordinates. It reads as
\begin{equation}
ds^{2}=-dt^{2}+dr^{2}+r^{2}d\theta ^{2}+\alpha ^{2}r^{2}\sin ^{2}\theta
d\varphi ^{2},  \label{21}
\end{equation}
with $0<r<\infty $, $0<\theta <\pi $, $0\leq \varphi \leq 2\pi $. In the
special case $\alpha =1$ we obtain the Minkowski space in spherical
coordinates.

In this case, the time-independent Schr\"{o}dinger equation is
\begin{eqnarray}
&&\left. -\frac{\hbar ^{2}}{2\mu r^{2}}\left[ 2r\frac{\partial }{\partial
r^{2}}+r^{2}\frac{\partial ^{2}}{\partial r^{2}}+\cot \theta \frac{\partial
}{\partial \theta }+\frac{\partial ^{2}}{\partial \theta ^{2}}+\frac{1}{%
\alpha ^{2}\sin ^{2}\theta }\frac{\partial ^{2}}{\partial \varphi ^{2}}%
\right] \psi \left( r,\theta ,\varphi \right) \right.  \nonumber \\
&&\left. V(r)\psi \left( r,\theta ,\varphi \right) =E\psi \left( r,\theta
,\varphi \right) \right. .  \label{22}
\end{eqnarray}
This partial differential equation can be solved by
finding separated solutions of the form

\begin{equation}
\psi (r,\theta ,\varphi )=\frac{1}{\sqrt{2\pi }}e^{im\varphi }R(r)\Theta
(\theta ),  \label{23}
\end{equation}
for which $m=0, \;\pm 1, \; \pm 2 ...\; .$
When we substitute Eq. (\ref{23}) into Eq. (\ref{22}) we thus obtain
\begin{eqnarray}
&&\left[-\frac{\hbar ^{2}}{2\mu } \frac{\partial }{\partial r}
\left( r^2 \frac{\partial}{\partial r} \right)
+r^{2}
V(r) +\lambda%
\right] R(r) =ER(r)
\label{you}
\end{eqnarray}

\begin{eqnarray}
&&\frac{\hbar ^{2}}{2\mu }\left[ \frac{1}{\sin \theta} \frac{\partial
}{\partial \theta } \left( \sin \theta \frac{\partial}
{\partial \theta} \right)-\frac{m^{2}}{%
\alpha ^{2}\sin^2\theta} \right]\Theta(\theta)+\lambda \Theta(\theta) =0,
\label{24}
\end{eqnarray}
indicating a neat separation of the variables $ r$
and $ \theta$, with $ \lambda$ being the separation constant.

If we change the function $R(r)$ by $u(r)$, introducing
\begin{equation}
u(r)=rR(r),  \label{26}
\end{equation}
we find that $u(r)$ obeys the radial equation
\begin{equation}
-\frac{\hbar ^{2}}{2\mu }\frac{\partial ^{2}u(r)}{\partial r^{2}}%
+V(r)_{eff}u(r)=Eu(r),  \label{27}
\end{equation}
where $V_{eff}$ is the effective potential experienced by the particle and
given by
\begin{equation}
V(r)_{eff}=V(r)+\frac{\lambda }{r^{2}}.  \label{28}
\end{equation}

Equation (\ref{27}) is identical in form with the one-dimensional
 Schr\"{o}dinger equation for a potential $V(r)$  with the
addition of the term $ \frac{\lambda}{r^2}$ to the potential energy.

Now, we turn to Eq. (\ref{24}).
By a change of variables $\xi =\cos \theta ;$ $\xi \in \left[ -1,1\right] ;$
$\theta \in \left[ 0,\pi \right] $, and
 introducing $F(\xi )=\Theta (\theta ),$ Eq. (\ref{24}) is transformed into
\begin{equation}
\left( 1-\xi ^{2}\right) \frac{d^{2}F(\xi )}{d\xi ^{2}}-2\xi \frac{dF(\xi )}{%
d\xi }-\left( \frac{2\mu \lambda }{\hbar ^{2}}+\frac{m^{2}}{\alpha
^{2}\left( 1-\xi ^{2}\right) }\right) F(\xi )=0.  \label{32}
\end{equation}
Equation (\ref{32}) is a kind of associated Legendre equation
 with a dependence on a parameter $\alpha$, which lies between
$ 0$ and $1 $. Let us call it then generalized associated Legendre
 equation. It
becomes an equation with eigenvalue $\frac{2\mu \lambda }{\hbar ^{2}}$, if
we demand that the solution be finite at the singular points $\xi =\pm 1.$

A convenient method to obtain the solutions of Eq. (\ref{32}) is by the
investigation of its behaviour at the singular points $\xi =\pm 1$. To do
this let us translate the origin to $\xi =1$ and introduce a new variable $%
z=1-\xi $.

Then, Eq. (\ref{32}) turns into
\begin{equation}
z\left( 2-z\right) \frac{d^{2}F(z)}{dz^{2}}+2\left( 1-z\right) \frac{dF(z)}{%
dz}-\left( \frac{2\mu \lambda }{\hbar ^{2}}+\frac{m^{2}}{\alpha ^{2}z\left(
2-z\right) }\right) F(z)=0.  \label{33}
\end{equation}
The solution of Eq. ($\ref{33}$) can be expanded in a power series
\begin{equation}
F(z)=z^{s}\sum_{k=0}^{\infty }a_{k}z^{k}.  \label{34}
\end{equation}
Substituting Eq. (\ref{34}) into (\ref{33}) we get the
indicial equation
\begin{equation}
a_{0}\left( s(s-1)4z^{s}+4sz^{s}-\frac{m^{2}}{\alpha ^{2}}z^{s}\right) =0,
\label{35}
\end{equation}
which implies that
\[
s=\frac{m}{2\alpha },
\]
and therefore
\begin{equation}
F(\xi )=\left( 1-\xi \right) ^{\frac{m}{2\alpha }}f(\xi ),  \label{36}
\end{equation}
where $f(\xi )$ is an analytic function which does not vanish at $\xi =1$.

In order to investigate the behaviour of $F(\xi )$ in the neighborhood of $%
\xi =-1$ let us introduce the following substitution $z=1+\xi $. Then we get
that $F(\xi )=\left( 1+\xi \right) ^{\frac{m}{2\alpha }}g(\xi )$, and
therefore, the accepted solution may have the form
\begin{equation}
F(\xi )=\left( 1-\xi ^{2}\right) ^{\frac{m}{2\alpha }}G(\xi ),  \label{37}
\end{equation}
where $G(\xi)$ is an analytic function in all space, except
when $z\rightarrow \infty ,$ and
is different from zero for $\xi =\pm 1.$

From Eqs. ($\ref{32}$) and (\ref{37}), we get
\begin{equation}
\left( 1-\xi ^2\right) \frac{d^2G(\xi )}{d\xi ^2}-2(m_{(\alpha)}+1)\xi \frac{%
dG(\xi )}{d\xi }-\left( m_{(\alpha)}^2+m_{(\alpha)}+\bar{\lambda}\right)
 G(\xi )=0,
\label{38}
\end{equation}
where
\[
m_{(\alpha)}\equiv\frac m\alpha ;\text{ }\bar{\lambda}=
 \frac{2\mu \lambda }{\hbar ^2}.
\]
The substitution of Eq. (\ref{37}) into Eq. (\ref{38})
 yields the recursion relation
\begin{equation}
a_{n+2}=\frac{n\left( n-1\right) +2\left( m_{(\alpha)}+1\right)
 n+\bar{\lambda}+%
m_{(\alpha)}\left( m_{(\alpha)}+1\right) }{\left( n+1\right)
 \left( n+2\right) }a_n,
\label{39}
\end{equation}
where $n$ is an integer $\geq 0$. In order to acceptable
eigenfunctions, the series must terminate at some finite
value of $n$. According to Eq. (\ref{39}), this
will happen if $\bar{\lambda}$ has the value
\begin{equation}
\left.  \bar{\lambda}=l_{(\alpha )}\left( l_{(\alpha )}+1\right)
\right. ,  \label{40}
\end{equation}
where
\[
l_{(\alpha )}= m_{(\alpha)} +n.
\]
Therefore, Eq. (\ref{32}) turns into
\begin{equation}
\left( 1-\xi ^2\right) \frac{d^2F(\xi )}{d\xi ^2}-2\xi \frac{dF(\xi )}{d\xi }%
-l_{(\alpha )}(l_{(\alpha )}+1)F(\xi )-\frac{m^2}{\alpha ^2(1-\xi ^2)}F(\xi
)=0,  \label{41}
\end{equation}
whose solutions are the generalized associated Legendre functions
\begin{equation}
F_{l_{(\alpha )}}^{m_{(\alpha )}}(\xi )(\xi )=P_{l_{(\alpha )}}^{m_{(\alpha
)}}(\xi )=\frac 1{2^{l_{(\alpha )}}l_{(\alpha )}!}\left( 1-\xi ^2\right) ^{%
\frac{m_{(\alpha )}}2}\frac{d^{m_{(\alpha )}+l_{(\alpha )}}}{d\xi
^{m_{(\alpha )}+l_{(\alpha )}}}\left[ (\xi ^2-1)^{l_{(\alpha )}}\right] .
\label{42}
\end{equation}

Now, we turn to Eq. (\ref{27}), substitute the expression
for $ \bar {\lambda}$ given by Eq. (\ref{40}) and consider $V(r)=-%
\frac kr$. We thus obtain
\begin{equation}
\frac{d^2u(r)}{du^2}+2k\frac{\mu u(r)}{r\hbar ^2}-\bar{\beta}^2u(r)-\frac 1{%
r^2}\left[ l_{(\alpha )}(l_{(\alpha )}+1)\right] u(r)=0,  \label{43}
\end{equation}
where
\begin{equation}
\bar{\beta}^2=-\frac{2\mu E_{n_r}}{\hbar ^2};\text{ }E_{n_r}<0.  \label{g}
\end{equation}
Equation (\ref{43}) is a confluent hypergeometric equation whose solution is
given by
\begin{equation}
u(r)=_1F_1\left( l_{(\alpha )}+1-\frac{k^2\mu }{\bar{\beta}\hbar ^2}%
,2+2l_{(\alpha )};2\bar{\beta}r\right) .  \label{44}
\end{equation}
This function is divergent, unless
\begin{equation}
1+l_{(\alpha )}-\frac{\mu k^2}{\bar{\beta}\hbar ^2}=-n_r;\text{ }n_r=0,1,2...%
\text{ .}  \label{45}
\end{equation}
Then, from the previous condition we find the energy eigenvalues
\begin{equation}
E_{n_r}=-\frac{\mu k^2}{2\hbar ^2}\left[ l_{(\alpha )}+n_r^{\prime }\right]
^{-2};\text{ }n_r^{\prime }=1,2,3...\text{,}  \label{46}
\end{equation}
where $n_r^{\prime }=1+n_r$. Note that the energy levels become more and
more spaced as $\alpha $ tends to $1$, which means that the shift in the
energy levels due to the presence of the cosmic string increases with the
increasing of the angular deficit.

From the expression for energy given by Eq. (\ref{46}) we can notice that the
levels without a $z$-component of the angular momentum are not shifted
relative to the Minkowski case. Except these levels, all the other ones are
degenerated.

In this present case the energy of the particle in the presence of the
cosmic string increases as compared with the flat spacetime value. This
increasing is of about $4\times 10^{-3}\%$ for the first two energy levels.

\vskip 2.0 cm

\centerline{\bf {VI. Kratzer potential in the spacetime of a global monopole}%
}

The solution corresponding to a global monopole in a $O(3)$ broken symmetry
model has been investigated by Barriola and Vilenkin\cite{8}.

Far away from the global monopole core we can neglect the mass term and as a
consequence the main effects are produced by the  deficit solid angle. The
respective metric in Einstein$^{\prime }$s theory of gravity can be written
as\cite{8}
\begin{equation}
ds^2=-dt^2+dr^2+b^2r^2\left( d\theta ^2+\sin ^2\theta d\varphi ^2\right) ,
\label{c3}
\end{equation}
where $b^2=1-8\pi G\eta ^2$, $\eta $ being the energy scale of symmetry
breaking.

Now, let us consider a particle interacting with a Kratzer potential and
placed in the background spacetime given by metric (\ref{c3}).

The Kratzer potential has the form\cite{15}
\begin{equation}
V(r)=-2D\left( \frac Ar-\frac 12\frac{A^2}{r^2}\right) ,  \label{a}
\end{equation}
where $A$ and $D$ are positive constants.

In order to determine the energy spectrum let us write the Schr\"{o}dinger
equation in the background spacetime of a global monopole given by line
element(\ref{c3}). In this case the Schr\"{o}dinger equation can thus be
reduced to
\begin{equation}
-\frac{\hbar ^2}{2\mu b^2r^2}\left[ 2rb^2\frac \partial {\partial r}+b^2r^2%
\frac{\partial ^2}{\partial r^2}-{\bf L}^2-2D\left( \frac Ar-\frac 12\frac{%
A^2}{r^2}\right) \right] \psi ({\bf r})=E\psi ({\bf r}),  \label{b}
\end{equation}
where ${\bf L}$ is the usual orbital angular momentum operator. We begin by
using the standard procedure for solving Eq. (\ref{b}) which consists in the
separation of the eigenfunctions as

\begin{equation}
\psi _{m,l}({\bf r})=R_l(r)Y_l^m(\theta ,\varphi ).  \label{c}
\end{equation}
Substitution of Eq. (\ref{c}) into Eq. (\ref{b}) leads to
\begin{equation}
-\frac{\hbar ^2}{2\mu }\frac{d^2g_l(r)}{dr^2}-2D\left( \frac Ar-\frac 12%
\frac{A^2}{r^2}\right) g_l(r)+\frac{\hbar ^2}{2\mu }\frac{l(l+1)}{b^2r^2}%
g_l(r)=Eg_l(r),  \label{d}
\end{equation}
where $g_l(r)=rR_l(r).$

Analyzing Eq. (\ref{d}) when $r\rightarrow 0$ and $r\rightarrow \infty $ we
find that its solution can be written as
\begin{equation}
g_l(r)=r^{\lambda _l}e^{-\bar{\beta}r}F_l(r),  \label{e}
\end{equation}
where
\begin{equation}
\left. \lambda _l=\frac 12+\frac 12\sqrt{1+4\left( \frac{2\mu DA^2}{\hbar ^2}%
+\frac{l\left( l+1\right) }{b^2}\right) }\text{  }\right.   \label{f}
\end{equation}
and
\begin{equation}
\bar{\beta}=-\frac{2\mu E}{\hbar ^2},\text{ }E<0.  \label{fg}
\end{equation}
Substituting Eq. (\ref{e}) into Eq. (\ref{d}) and making use of Eqs. (\ref{g}%
) and (\ref{f}) we obtain the equation for $F(z)$
\begin{equation}
z\frac{d^2F(z)}{dz^2}+(2\lambda _l-z)\frac{dF(z)}{dz}-\left( \lambda _l-%
\frac{2mAD}{\bar{\beta}\hbar ^2}\right) F(z)=0,  \label{h}
\end{equation}
where $z=2\bar{\beta}r$.

The solution of this differential equation is the confluent
 hypergeometric function $_1F_1\left( \lambda _l-
\frac{\gamma ^2}{\bar{\beta}A}%
,2\lambda _l\text{ ; }2\bar{\beta}r\right) $, where $\gamma ^2
=\frac{2\mu DA^2%
}{\hbar ^2}.$

Therefore, the solution for the radial function $g_l(r)$ is given by
\begin{equation}
g_l(r)=r^{\lambda _l}e^{-\bar{\beta}r}{}_1F_1\left( \lambda _l-\frac{\gamma
^2}{\bar{\beta}A},2\lambda _l\text{ ; }2\bar{\beta}r\right) .  \label{i}
\end{equation}

In order to make $g_l(r)$ vanishes for $r\rightarrow \infty $, the confluent
hypergeometric function may increase not faster than some power of $r$, that
is, the function must be a polynomial. To fulfill this condition we must have
\begin{equation}
\lambda _l-\frac{\gamma ^2}{\bar{\beta}A}=-\bar{n}_r,\text{ }\bar{n}%
_r=0,1,2,...\text{ },  \label{y}
\end{equation}
which implies that the eigenvalues are
\begin{equation}
E_{l,\bar{n}_r}=-\frac{\hbar ^2}{2\mu A^2}\gamma ^4\left( \bar{n}_r+\frac 12+%
\sqrt{\frac 14+\frac{l\left( l+1\right) }{b^2}+\gamma ^2}\right) ^{-2}
\label{l}
\end{equation}

Let us consider the parameter $\gamma \gg 1$, which is a situation valid for
the majority of cases. Then we may expand Eq. (\ref{l}) into powers of $%
\frac 1\gamma$, and obtain the approximate result

\begin{eqnarray}
E_{l,\bar{n}_r} &\cong &D\left( -1+2\frac{\left( \bar{n}_r+\frac 12\right) }%
\gamma +\frac{\left( \frac 14+l\frac{\left( l+1\right) }{b^2}\right) }{%
\gamma ^2}\right.   \nonumber \\
&&\left. -3\frac{\left( \bar{n}_r+\frac 12\right) }{\gamma ^2}^2-3\frac{%
\left( \bar{n}_r+\frac 12\right) \left( \frac 14+l\frac{\left( l+1\right) }{%
b^2}\right) }{\gamma ^3}\right) .  \label{o}
\end{eqnarray}
Introducing the momentum of inertia defined by
\begin{equation}
\Theta =\mu A^2,  \label{p}
\end{equation}
and using the classical frequency for small harmonic vibrations
 which is given by

\begin{equation}
w=\sqrt{\frac{2D}{A^2\mu }},  \label{g1}
\end{equation}
we thus find that  Eq. (\ref{o}) transforms into

\begin{eqnarray}
E_{l,\bar{n}_r} &\cong &-\frac{w^2\Theta }2+\left( \bar{n}_r+\frac 12\right)
\hbar w+\frac{\hbar ^2}{2\Theta }\left( \frac 14+l\frac{\left( l+1\right) }{%
b^2}\right)   \nonumber \\
&&-3\left( \bar{n}_r+\frac 12\right) ^2\frac{\hbar ^2}{2\Theta }-\frac 32%
\left( \bar{n}_r+\frac 12\right) \left( \frac 14+l\frac{\left( l+1\right) }{%
b^2}\right) \frac{\hbar ^3}{w\Theta ^2}.  \label{q}
\end{eqnarray}
From Eq. (\ref{q}) we can get the energy of dissociation of a molecule by
taking $\bar{n}_r=0$ and $l=0$, in which case there is no dependence on
the presence of the monopole. This result mean that the energy of
dissociation is no affected by the presence of the global monopole.

From the above results given by Eqs. (\ref{l}) and (\ref{q}), we can see
that when $b=1$ we recover the well-known result corresponding to a particle
submitted to the Kratzer potential\cite{15} in Minkowski spacetime as expected.

It is worthy noticing from expression for the  energy given by Eq. (\ref{l})
that even in the case in which the $z-$component of the angular momentum
vanishes the energy level is shifted relative to the Minkowski case.

As an estimation of the effect of the global monopole on the energy
spectrum, let us consider a stable global monopole configuration for which $%
\eta =0.19m_p$, where $m_p$ is Planck mass. In this situation the shift in
the energy spectrum between the first two levels in this background
spacetime decreases of about $350\%$ as compared with the Minkowski
spacetime. For symmetry breaking at grand unification scale, the typical
value of 8$\pi G\eta ^2$ is around $10^{-6}$ and in this case decreasing in
the energy shift is of about $1\%$.

\vskip 2.0 cm

\centerline{\bf{V. Morse potential in the spacetime of a global monopole }}

Now, let us consider the Morse potential\cite{16} which is given by

\begin{equation}
V(r)=D\left( e^{2\beta x}-2e^{-\beta x}\right) ;\text{ }x=\frac{r-r_0}{r_0};%
\text{ }\beta >0.  \label{47}
\end{equation}
In order to determine of the energy levels in a simply way, let us consider
a particular situation that corresponds to low-energy vibration. Hence, in this
case, we can expand this potential around $r=r_0$ (or $x=0$)
and obtain the approximate result

\begin{eqnarray}
V(r) &=&D(-1+\beta ^2x^2+...)  \nonumber \\
&\cong &-D+D\frac{\beta ^2}{r_0^2}\left( r-r_0\right) ^2.  \label{48}
\end{eqnarray}
Introducing a frequency $w$ defined by
\begin{equation}
w^2=\frac{2D\beta ^2}{Mr_0^2},  \label{tv}
\end{equation}
we can write Eq. (\ref{48}) as
\begin{equation}
V(r^{\prime })\cong -D+\frac 12Mw^2r^{\prime 2}\text{,}  \label{49}
\end{equation}
where  $r^{\prime }=r-r_0$.

This potential corresponds to the one of an  isotropic harmonic
oscillator minus
a constant. Substituting this potential into the radial part
 the Schr\"{o}dinger, we get
\begin{equation}
\frac{d^2g(r^{\prime })}{dr^{\prime 2}}-\frac{2M^2}{\hbar ^2}V(r^{\prime
})g(r^{\prime })-l\frac{\left( l+1\right) }{b^2r^{\prime 2}}g(r^{\prime })+%
\frac{2M}{\hbar ^2}E_Mg(r^{\prime })=0,  \label{52}
\end{equation}
where
\[
g(r^{\prime })=r^{\prime }R(r^{\prime }).
\]
The solution of Eq. (\ref{52}) is given by
\begin{equation}
g(r^{\prime })=\exp \left( -\frac 12M\frac w\hbar r^{\prime 2}\right)
r^{\prime \frac 12+\frac 12\sqrt{1+\frac 4{b^2}l(l+1)}}F(r^{\prime }).
\label{53}
\end{equation}
Substituting Eq. ($\ref{53}$) into Eq. (\ref{52}) we obtain
\begin{equation}
r^{\prime }\frac{d^2F(r^{\prime })}{dr^{\prime 2}}+\left[ c-\frac{2Mw}\hbar
r^2\right] \frac{dF(r^{\prime })}{dr^{\prime }}+\left[ P-\frac{2Mw}\hbar
\right] r^{\prime }F(r^{\prime })=0,  \label{54}
\end{equation}
with
\[
c=1+\sqrt{1+\frac 4{b^2}l\left( l+1\right) }\text{ and  }P=\frac{2M}{%
\hbar ^2}(E_M+D)-\frac{Mw}\hbar \sqrt{1+\frac 4{b^2}l\left( l+1\right) }
\]
Performing the change of variables
\begin{equation}
Mwr^{\prime 2}=\rho ,  \label{tv1}
\end{equation}
and defining  $\frac \rho \hbar =x,$ we can write Eq. (\ref{54}) as
\begin{equation}
x\frac{d^2F(x)}{dx^2}+\left[ \frac c2-x\right] \frac{dF(x)}{dx}+\left[ \frac{%
P\hbar }{4Mw}-\frac 12\right] F(x)=0,  \label{56}
\end{equation}
whose solution is
\begin{equation}
F(x)=_1F_1\left( \frac 12-\frac{\hbar P}{4Mw},\frac c2;x\right) ,
\label{val}
\end{equation}
or, in terms of variable $r^{\prime }$, we get
\begin{equation}
F(r^{\prime })=_1F_1\left( \frac 12-\frac{1}{2\hbar w}(E_M+D)+%
\frac 14\sqrt{1+\frac 4{b^2}l(l+1)},\frac 12+\frac 12\sqrt{\frac 4{b^2}l(l+1)%
};\frac{Mwr^{\prime 2}}\hbar \right) .  \label{57}
\end{equation}
Note that this solution is divergent, unless the condition
\begin{equation}
\frac 12-\frac{1}{2\hbar w}(E_M+D)+\frac 14\sqrt{1+\frac 4{b^2}%
l(l+1)}=-n_M;\text{ }n_M=0,1,2,3,...,  \label{58}
\end{equation}
is fulfilled.

From Eq. (\ref{58}) we find the energy levels
\begin{equation}
E_M=\hbar w\left[ N_{lM}+\frac 32\right] -D,  \label{59}
\end{equation}
where
\begin{equation}
N_{lM}=\frac 12\left( \sqrt{1+\frac 4{b^2}l\left( l+1\right) }-1\right)
+2n_M.  \label{60}
\end{equation}

Now, let us consider an additional term in the Morse potential which
corresponds to the rotation energy. Since, we can write $V(r)$ as
\begin{equation}
V(r)=D\left( e^{-2\beta x}-e^{-\beta x}\right) +V^{\prime }(r),  \label{60}
\end{equation}
where
\begin{equation}
V^{\prime }(r)=\frac{\hbar ^2}{2M}l\frac{\left( l+1\right) }{b^2r^2}=l\frac{%
\left( l+1\right) }{b^2\gamma ^2}\frac D{\left( x+1\right) ^2};\text{ }%
\gamma ^2=\frac{2MDr_0^2}{\hbar ^2}.  \label{61}
\end{equation}
Hence we can expand $V^{\prime }(r)$ and get the result
\begin{equation}
V^{\prime }(r)\cong l\frac{\left( l+1\right) }{b^2\gamma ^2}D\left(
c_0+c_1e^{-\beta x}+c_2e^{-2\beta x}\right) ,  \label{62}
\end{equation}
for $\left| x\right| \ll 1$, where the coefficients $c_0$, $c_1$ and $c_2$ are
given by
\begin{equation}
c_0=1-\frac 3\beta -\frac 3{\beta ^2};\text{ }c_1=\frac 4\beta -\frac 6{%
\beta ^2};\text{ }c_2=-\frac 1\beta +\frac 3{\beta ^2}.  \label{geusa}
\end{equation}

Substituting Eq. (\ref{62}) into Eq. (\ref{60}) and then into Eq. (\ref{52}%
), we get
\begin{equation}
\frac{d^2g(x)}{dx^2}+\left[ -\beta _1^2+2e^{-\beta x}\gamma _1^2-e^{-2\beta
x}\gamma _2^2\right] g\left( x\right) =0  \label{65}
\end{equation}
where
\begin{equation}
\beta _1^2=l\frac{\left( l+1\right) }{b^2}c_0-2\frac{ME_Mr_0^2}\hbar ;\text{
}E_M<0.  \label{66}
\end{equation}
\begin{equation}
\gamma _1^2=\gamma ^2-\frac{c_1}2l\frac{\left( l+1\right) }{b^2}.  \label{67}
\end{equation}
\begin{equation}
\gamma _2^2=\gamma ^2+c_2l\frac{\left( l+1\right) }{b^2}.  \label{68}
\end{equation}

Now, let us introduce the following change of variable
\begin{equation}
y=\xi e^{-\beta x},  \label{qw}
\end{equation}
and a function $R(y)$ such that
\begin{equation}
g(y)=\frac{R(y)}{\sqrt{y}}.  \label{vv}
\end{equation}
The equation for $R(y)$ is
\begin{equation}
y^2\frac{d^2R(y)}{dy^2}+\frac{\gamma _1^2}{\gamma _2\beta }yR(y)+\left[
\frac 14-\frac 14y^2-\frac{\beta _1^2}{\beta ^2}\right] R(y)=0,  \label{70}
\end{equation}
whose solution is given by
\begin{equation}
R(y)=\exp \left( -\frac 12y\right) y^{\frac 12+\left| \frac{\beta _1}\beta
\right| }F(y),  \label{71}
\end{equation}
with $F(y)$ obeying the Kummer equation
\begin{equation}
y\frac{d^2F(y)}{dy^2}+\left( 1+2\left| \frac{\beta _1}\beta \right|
-y\right) \frac{dF(y)}{dy}-\left( \frac 12+\left| \frac{\beta _1}\beta
\right| -\frac{\gamma _1^2}{\gamma _2\beta }\right) F(y)=0,  \label{72}
\end{equation}
whose solution is the Kummer function
\begin{equation}
F(y)=M\left( \frac 12+\left| \frac{\beta _1}\beta \right| -\frac{\gamma _1^2%
}{\gamma _2\beta },1+2\left| \frac{\beta _1}\beta \right| ;y\right) .
\label{73}
\end{equation}
This solution diverges, unless we have
\begin{equation}
\frac 12+\left| \frac{\beta _1}\beta \right| -\frac{\gamma _1^2}{\gamma
_2\beta }=-n;\text{ }n=0,1,2,...\text{ }  \label{74}
\end{equation}

This relation leads thus to the energy levels
\begin{eqnarray}
E_M &=&\frac{\hbar ^2}{2Mr_0^2}\left[ -\gamma ^2+2\gamma \beta \left( n+%
\frac 12\right) -\beta ^2\left( n+\frac 12\right) ^2-l^2\left( l+1\right) ^2%
\frac{\left( c_1+c_2\right) ^2}{4b^2\gamma ^2}\right.   \nonumber \\
&&\left. +l\frac{\left( l+1\right) }{b^2}\left( c_0+c_1+c_2\right) -l\frac{%
\left( l+1\right) \beta }{b^2}\left( c_1+c_2\right) \left( n+\frac 12\right)
\right] .  \label{76}
\end{eqnarray}
 which was obtained by use of the approximate result
\begin{eqnarray*}
\frac{\gamma _1^2}{\gamma _2} &=&\left[ \gamma ^2-\frac{c_1}2l\frac{\left(
l+1\right) }{b^2}\right] \left[ \gamma ^2-\frac{c_2}2l\frac{\left(
l+1\right) }{b^2}\right] ^{-\frac 12} \\
&\approx &\gamma \left( 1-l\frac{\left( l+1\right) }{2b^2\gamma ^2}\left(
c_1+c_2\right) \right) ,
\end{eqnarray*}
Finally, substituting the values of the coefficients $c_0$, $c_1$and $c_2$
given by Eq. (\ref{geusa}) into Eq. (\ref{76}) we find
\begin{eqnarray}
E_M &=&\frac{\hbar ^2}{2Mr_0^2}\left[ -\gamma ^2+2\gamma \beta \left( n+%
\frac 12\right) -\beta ^2\left( n+\frac 12\right) ^2+l\frac{\left(
l+1\right) }{b^2}\right.   \nonumber \\
&&\left. -\frac{3\left( \beta -1\right) }{\beta \gamma }\left( n+\frac 12%
\right) l\frac{\left( l+1\right) }{b^2}-\frac{9\left( \beta -1\right) ^2}{%
4\gamma ^2}l^2\frac{\left( l+1\right) ^2}{b^2}\right] .  \label{77}
\end{eqnarray}
which reduces to the result of flat spacetime when $b=1$ as expected.

In the case of a particle interacting with a  Morse potential in the
background spacetime of a global monopole we can consider two different
situations. In the first, only the vibration energy is taken
into account and in this case the decreasing in the energy goes from $%
3\times 10^{-5}\%$ to $32\%$ in the cases of GUT and stable monopoles,
respectively, as compared to the flat spacetime background. In the second
situation, we consider also the corrections due to the rotational energy and in
this case we observe that there is a decrease in the energy that goes from $%
8\times 10^{-4}\%$ to $85\%$ compared to Minkowski spacetime case, for GUT
and stable monopoles, respectively.

\section{FINAL REMARKS}

In the spacetime of a cosmic string we studied the behaviour of a particle
interacting with an harmonic oscillator and a Coulomb potentials. The quantum
dynamics of such a single particle depends on the nontrivial topological
features of the cosmic string spacetime. The presence of the defect shift
the energy levels in both cases relative to Minkowski spacetime case. These
gravitational effects can be understood as a kind of Aharonov-Bohm effect
for bound states and as a consequence extends the Aharonov-Bohm effect
beyond electromagnetic theory to the gravitational field.

In the case of the molecular potentials of Kratzer and Morse in the presence
of a global monopole the shifts in the energy levels are due to the combined
effects of the curvature and the nontrivial topology determined by the
deficit solid angle corresponding to this spacetime.

The numerical estimations concerning the energy shifts for different
potentials in the backgrounds of a cosmic string and a global monopole are
all measurable, in principle. An important application of the results
concerning the modifications in the energy spectra could be found in the
astrophysical context in which these modifications enters the interpretation
of the spectroscopical data and these could be used as a probe of the
presence of a cosmic string or a global monopole in the universe.

Although it seems difficult to provide verification of the effects
considered with current experimental detectability, we believe that
investigation of systems in which both quantum effects and gravitational
effects associated with curvature and topology come into play is important.

It is interesting to draw the attention to the fact that
 the presence of cosmic string in the cases considered
increases the energy which is in contrast to the case of
 the presence of a monopole in which situation the energy
 decreases as compared with the flat Minkowski spacetime background.

It is worthy commenting that the study of a quantum system in a nontrivial
gravitational background like the ones corresponding to a cosmic string and
to a global monopole may shed some light on the problems of combining
quantum mechanics and general relativity, in situation in which
 nontrivial topological aspects of the background spacetime
 under consideration a represent.

\section*{Acknowledgments}

We acknowledge Conselho Nacional de Desenvolvimento Cient\'{\i }fico e
Tecnol\'{o}gico for partial financial support.

\end{document}